\documentclass[runningheads]{llncs}
\usepackage[T1]{fontenc}
\usepackage{graphicx}
\usepackage{xcolor}

\usepackage{hyperref}
\usepackage{cleveref}
\crefname{lstlisting}{listing}{listings}
\Crefname{lstlisting}{Listing}{Listings}

\usepackage{listings}
\lstdefinelanguage{Dafny}{
  morekeywords={
    method, function, lemma, ghost, var, const, new, class,
    requires, ensures, decreases, invariant, modifies, predicate,
    while, if, else, return, returns, type, datatype, then,
    forall, exists, match, case, import, module,
    assert, assume, calc, label, this, null, fresh, old
  },
  sensitive=true,
  morecomment=[l]{//},
  morecomment=[s]{/*}{*/},
  morestring=[b]",
}

\lstdefinelanguage{json}{
    morekeywords={
        user, system
    },
    basicstyle=\normalfont\ttfamily,
    numbers=left,
    numberstyle=\scriptsize,
    breaklines=true,
    frame=lines,
    backgroundcolor=\color{white},
    showstringspaces=false,
    string=[db]{"},
    stringstyle=\color{green!50!black},
    morestring=[s][\color{green!50!black}]{\ \ "}{"},
    comment=[l]{\#},
    keywordstyle=\color{blue},
    keywords={true,false,null},
    literate=
     *{0}{{{\color{red}0}}}{1}
      {1}{{{\color{red}1}}}{1}
      {2}{{{\color{red}2}}}{1}
      {3}{{{\color{red}3}}}{1}
      {4}{{{\color{red}4}}}{1}
      {5}{{{\color{red}5}}}{1}
      {6}{{{\color{red}6}}}{1}
      {7}{{{\color{red}7}}}{1}
      {8}{{{\color{red}8}}}{1}
      {9}{{{\color{red}9}}}{1}
      {.}{{{\color{red}.}}}{1}
      {:}{{{\color{gray}{:}}}}{1}
      {,}{{{\color{gray}{,}}}}{1}
      {\{}{{{\color{gray}{\{}}}}{1}
      {\}}{{{\color{gray}{\}}}}}{1}
      {[}{{{\color{gray}{[}}}}{1}
      {]}{{{\color{gray}{]}}}}{1},
}

\begin{document}
\title{Specification-Guided Repair of Arithmetic Errors in Dafny Programs using LLMs}
\titlerunning{Specification-Guided Repair of Dafny Programs using LLMs}
\author{Valentina Wu\inst{1}\orcidID{0009-0006-2472-8524} \and
Alexandra Mendes\inst{2}\orcidID{0000-0001-8060-5920} \and
Alexandre Abreu\inst{2}\orcidID{0000-0003-4198-3181}}
\authorrunning{V. Wu et al.}
\institute{
Faculdade de Engenharia,Universidade do Porto, Porto, Portugal\\
\email{up201907483@up.pt}
\and
INESC TEC, Faculdade de Engenharia, Universidade do Porto, Porto, Portugal\\
\email{alexandra@archimendes.com,alexandre.filho@fe.up.pt}
}
\maketitle              %
\begin{abstract}

Debugging and repairing faults when programs fail to formally verify can be complex and time-consuming. %
Automated Program Repair (APR) can ease this burden by automatically identifying and fixing faults. 
However, traditional APR techniques often rely on test suites for validation, but these may not capture all possible scenarios. In contrast, formal specifications provide strong correctness criteria, enabling more effective automated repair.

In this paper, we present an APR tool for Dafny, a verification-aware programming language that uses formal specifications\,---\,including pre-conditions, post-conditions, and invariants\,---\,as oracles for fault localization and repair. Assuming the correctness of the specifications and focusing on arithmetic bugs, we localize faults through a series of steps, which include using Hoare logic to determine the state of each statement within the program, and applying %
Large Language Models (LLMs) to synthesize candidate fixes. The models considered are GPT-4o mini, Llama 3, Mistral 7B, and Llemma 7B.

We evaluate our approach using DafnyBench, a benchmark of real-world Dafny programs. Our tool achieves 89.7\% fault localization success rate and GPT-4o mini yields the highest repair success rate of 74.18\%. These results highlight the potential of combining formal reasoning with LLM-based program synthesis for automated program repair.

\end{abstract}
\section{Introduction}\label{sec:intro}
Software is essential in daily life, impacting communication, transportation, healthcare, and more. Despite careful development, bugs can cause unexpected behaviour or system failures, making their identification and repair both critical and time-consuming. Automated program repair (APR)~\cite{apr_17} 
aims to automatically identify bugs and generate fixes without direct human intervention, thus improving the efficiency of software development. 
However, APR methods often rely on weak correctness criteria, such as tests, which do not guarantee overall program correctness\,---\,a limitation that is particularly problematic in critical systems. %

Contract programming~\cite{designbycontract} improves software correctness by using formal specifications to define expected program behaviour through pre-conditions, post-conditions, and invariants. While effective, contract programming can be complex and resource-intensive, and manual proofs are often required for successful verification. Verification-aware languages, such as Dafny~\cite{leino2010dafny}, support contract programming and formal verification, enabling the detection of programs that do not satisfy their specifications. However, repairing programs when verification fails remains challenging. APR is therefore valuable in this context, especially given the availability of specifications that precisely define expected behaviour and provide strong correctness criteria for repair.

In this paper, we present an APR tool for Dafny that uses formal specifications as the oracle for repair. This process involves localizing faults using Hoare logic rules, and generating fix candidates using a Large Language Model (LLM)~\cite{llms}. The candidates are then verified against the Dafny specification to determine if they repair the program. Our main contributions are:
\begin{itemize}
    \item We propose an APR technique for Dafny that uses formal specifications as the oracle, eliminating reliance on test suites;
    \item We integrate fault localization based on Hoare logic with patch synthesis guided by LLMs;
    \item We implement a repair tool and evaluate it on DafnyBench, achieving 89.7\% fault localization success rate and 74.18\% repair success using GPT-4o mini;
    \item We provide a comparative analysis of GPT-4o mini\footnote[1]{\url{https://platform.openai.com/docs/models/gpt-4o-mini.}}, Llama 3\footnote[2]{\url{https://huggingface.co/meta-llama/Meta-Llama-3-8B}}, Mistral 7B\footnote[3]{\url{https://huggingface.co/mistralai/Mistral-7B-v0.1}}, and Llemma 7B\footnote[4]{\url{https://huggingface.co/EleutherAI/llemma_7b}}, and their performance in formal verification-driven repair.
\end{itemize}

The remainder of this paper presents the necessary background, our approach, and the obtained results. Our tool is available at \url{https://github.com/VeriFixer/APRepair-of-Arithmetic-Programs-in-Dafny-using-LLMs}. %

\section{Background}\label{sec:bg}
This section provides the necessary background on Dafny, Hoare logic, and LLMs, which underpin the proposed approach.

\paragraph{\bf The Dafny Language and Verifier.}

Dafny~\cite{leino_dafny_2014} is a programming language designed for formal software verification, using constructs for program specifications that define expected behaviour and allow mathematical proof of consistency between specifications and implementations.

The Dafny verifier relies on Boogie, an intermediate verification language, and Z3~\cite{de_moura_z3_2008}, a powerful theorem prover. Dafny code is first translated into Boogie, generating first-order verification conditions that Z3 validates. If valid, the program is confirmed as verified; if not, a counterexample is produced, leading to error messages that pinpoint issues in the code.

Dafny employs pre-conditions (\emph{requires}) and post-conditions (\emph{ensures}) to define specifications.
Loop invariants (\emph{invariant}) ensure conditions hold during iterations, while class invariants maintain object consistency at method entry and exit points. Dafny also supports ghost entities, which are used solely for verification. %
These constructs are exemplified in~\Cref{lst:dfy}.

\begin{lstlisting}[label={lst:dfy},caption={Example of a Dafny Method with Annotations}]
method example(n: int) returns (i : int)
  requires n >= 0 // property of the input, verified for each call
  ensures i == n // property of the output
{
  i := 0; // variable assignment
  while i < n
    invariant 0 <= i <= n // property that holds before and after each                            iteration of the loop
  {
    i := i + 1; // simple statement
  } // at this point, the post-condition is verified be true
}
\end{lstlisting}

\paragraph{\bf Hoare Logic.}
Hoare logic~\cite{hoare_axiomatic_1969} is a formal system with a set of logical rules used to reason about the correctness of computer programs. %
It employs deductive reasoning to prove properties of programs using Hoare Triples, denoted as $\{P\}\, Q \,\{R\}$, where $P$ represents pre-conditions, $Q$ is a sequence of program statements, and $R$ represents post-conditions. This syntax means that if $P$ holds before executing $Q$, $R$ will hold afterwards, if $Q$ terminates.

The reasoning for Hoare logic rules involves verifying post-conditions for assignments, sequencing statements, and evaluating both branches in conditionals. For reasoning about loops (\texttt{while} $B$ \texttt{do} $S$), correctness requires invariants, $I$, that act as both pre-conditions and post-conditions, %
where three assertions must be true: \emph{initialization}\,---\,the invariant must hold before the first iteration ($P \rightarrow I$); \emph{maintenance}\,---\,if the invariant holds and the loop condition is true, executing the loop body must preserve the invariant ($\{I \land B\}\,S\,\{I\}$); and \emph{termination}\,---\,when the loop exits ($B$ is false), the post-condition $R$ must hold  ($(I \land \neg B) \rightarrow R$).

\paragraph{\bf LLMs for Code Synthesis.}
LLMs are based on transformer architectures and use deep learning to perform a range of natural language processing tasks, such as text understanding, translation, and content generation. LLMs generate outputs by analyzing extensive text data through an attention mechanism emphasizing relevant input features.

Working with LLMs primarily involves two main approaches: fine-tuning and prompting~\cite{zhao_explainability_2024}. Fine-tuning involves initial pre-training on unlabeled data followed by their adaptation to specific labelled data to improve performance in particular domains. In the prompting approach, models use examples and task descriptions to enhance accuracy, using techniques like zero-shot prompting, where no examples are provided, or few-shot prompting, where a small number of examples are given. In this paper, we only use prompting.

\section{Related Work}\label{sec:rw}

This section reviews prior work in APR, fault localization, and using formal specifications and LLMs for program synthesis.

\paragraph{\bf Automated Program Repair.}

Automated Program Repair aims to enhance software development by automatically fixing program bugs~\cite{le_goues_automated_2019,le_goues_automatic_2021}, significantly reducing the time and effort required to debug and test software. %
The process involves identifying failure causes, generating patch candidates, and evaluating modifications to ensure the program passes all tests without introducing new bugs. Key steps include taking a buggy program and a test suite, locating the bug, generating potential patches, and validating them against the test suite.

Three main approaches to APR~\cite{le_goues_automated_2019} are:
\begin{itemize}
    \item \textbf{Heuristic Repair}: This employs a generate-and-test approach where patch candidates are created and validated based on the number of passing tests. Genetic Program Repair (GenProg)~\cite{le_goues_genprog_2012} %
    is a notable example that uses a fitness function to evaluate program variants;
    \item \textbf{Constraint-based Repair}: This approach constructs repair constraints that the patch must satisfy. Semfix~\cite{nguyen_semfix_2013} is an example tool that, using symbolic execution, constraint solving, and program synthesis to generate repairs, modifies faulty statements until the program passes all defined tests;
    \item \textbf{Learning-based Repair}: This method uses machine learning models trained on large datasets to generate patches automatically. The focus is often on producing realistic repairs that align with the structure and style of existing code~\cite{xia_less_2022}.
\end{itemize}

\paragraph{\bf Advanced Fault Localization Techniques.}

Traditional fault localization approaches~\cite{wong_survey_2016} include, among others, spectrum-based methods~\cite{abreu_accuracy_2007,wong_dstar_2014}, which correlate concrete execution traces with test outcomes, and mutation-based localization, which assess the effect of minor code changes on program behaviour. However, these approaches rely heavily on dynamic execution data. %

In contrast, static analysis techniques provide a way to localize faults without executing the code. One common static technique is slicing, which tries to remove lines irrelevant to the fault~\cite{wong_survey_2016}. We employ a static fault localization approach based on Hoare logic, using entailments between pre- and post-conditions to identify likely buggy statements, taking advantage of logic and the specification information available.

\paragraph{\bf Contract-Based APR.}

Traditional APR techniques often rely on test suites as the primary correctness oracle. While practical and widely adopted, this approach faces significant limitations. One major drawback is the inherent incompleteness of test suites, which may fail to fully capture a program's intended behaviour. As a result, APR methods that generate and validate patches against the same limited test set risk producing overfitting patches\,---\,fixes that pass all available tests but are semantically incorrect or incomplete~\cite{smith_is_2015}. This overfitting problem undermines the reliability of such patches, especially when the test coverage is low or the test inputs are unrepresentative of real-world usage.

In contrast, formal verification offers a more robust alternative by providing mathematical correctness guarantees concerning explicitly defined specifications~\cite{moy_testing_2013}. Rather than relying on examples, formal methods use logical reasoning to prove that a program adheres to its pre- and post-conditions. This approach can verify not just the absence of specific bugs but the correctness of entire classes of behaviours. However, the effectiveness of formal verification is contingent on the quality of the specifications themselves. Crafting specifications is a complex task that demands expertise and introduces its own potential for errors.

Several studies have proposed contract-based APR. Notably, previous work by Nguyen et al.~\cite{nguyen_automatic_2019} introduced a method that uses Hoare logic and program specifications to localize faults. Their approach involves computing the state at each program point via Hoare rules, generating logical entailments that must hold for the specification to be satisfied. Violated entailments indicate faulty code, which is then patched using linear expression templates with unknown coefficients. The system of constraints derived from the failed entailments is solved to infer the values of these coefficients, producing candidate repairs.

Other contract-driven APR approaches vary in bug localization techniques but similarly depend on specifications as oracles~\cite{konighofer_automated_2011,pei_code-based_2011,wei_automated_2010}, %
highlighting the utility of contracts in improving fault localization precision~\cite{pei_code-based_2011}. Furthermore, contract repair studies such as \cite{abreu_exploring_2023,pei_automatic_2014} focus on fixing the specification itself rather than the code. These methods generate tests dynamically to observe program behaviour, identify contract violations, and synthesize fixes by strengthening or weakening contract clauses to better align with observed behaviour.

Our approach assumes that the specification is correct and uses it as a reliable oracle for both fault localization and repair. We focus specifically on arithmetic bugs and use formal reasoning alongside LLMs to synthesize verified patches, further extending the applicability of contract-aware APR methods in verification-aware languages like Dafny.

\paragraph{\bf LLM Usage.}

Recent advances in LLMs, such as Codex, GPT-4, and Llama, have shown strong results in program repair, focusing on prompt engineering~\cite{xia_automated_2023,prenner_can_2022}. Tools like AlphaRepair~\cite{xia_less_2022} and CodeBERT~\cite{feng_codebert_2020} use trained models to predict fixes given buggy code. While these tools often lack formal guarantees, they demonstrate strong generalization in real-world settings.

Our work integrates LLMs into a formally grounded repair pipeline, where generated patches are only accepted if they pass formal verification. This hybrid approach combines the generative power of LLMs with the rigour of formal methods, producing both expressive and correct repairs.

Recent research has explored the application of LLMs to formal verification tasks as well, including work specific to the Dafny programming language. LLMs have been used to generate complete Dafny programs~\cite{misu_towards_2024,sun_clover_2024}, to synthesize loop invariants and assertions to support program correctness~\cite{loughridge_dafnybench_2024,mugnier_laurel_2024}, and to produce auxiliary lemmas that assist verification when Dafny's built-in verifier encounters complex reasoning challenges~\cite{silva_leveraging_2024}. The use of ChatGPT by students to solve formal verification exercises in Dafny has also been studied~\cite{carreira2025DafnyStudentStudy}. These works demonstrate the growing role of LLMs in enhancing developer productivity in verification-aware environments.
\section{Approach}\label{sec:appr}

\begin{figure}[htbp]
  \centering
    \includegraphics[width=\linewidth]{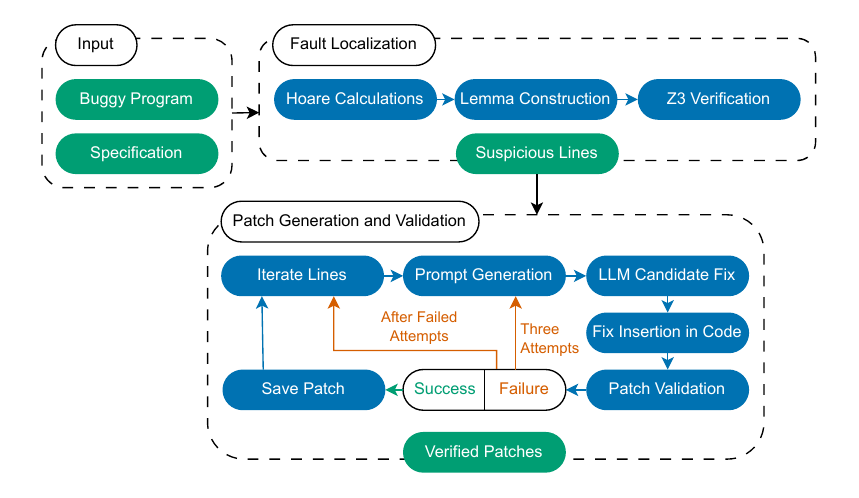}
  \caption{Overview of the Solution Pipeline}\label{fig:pipeline}
  \label{fig:method}
\end{figure}

Our approach to repairing arithmetic bugs in Dafny programs uses formal specifications as correctness oracles and LLMs for patch generation, as illustrated in~\Cref{fig:pipeline}. We assume that each program contains a single bug and that the formal specification is correct. Our approach consists of three main phases: \begin{enumerate}
    \item Fault localization using formal reasoning;
    \item Patch generation using LLMs;
    \item Patch validation using the Dafny verifier.
\end{enumerate}

The tool receives a buggy Dafny program and its corresponding specification. It first identifies a ranked list of suspicious lines using static analysis. For each suspicious line, it then queries an LLM to generate candidate fixes. Each candidate is inserted into the program and checked by the verifier. If verification fails, the model is queried again (up to three times per line). The process terminates when a verified patch is found or all options are exhausted.

\subsection{Fault Localization using Hoare Logic}

To identify the location of arithmetic bugs in Dafny programs, we employ a static analysis technique inspired by the method of Nguyen et al.~\cite{nguyen_automatic_2019}. Our approach relies on formal specifications provided by the developer and uses Hoare logic to analyze how each statement transforms the program state. The goal is to identify violations of expected logical entailments between pre- and post-conditions.

At each key control point\,---\,such as return statements and loop boundaries\,---\,we compute the expected post-state using Hoare rules. We then check whether the inferred program state implies the formal specification. If the entailment fails, the associated statement is marked as suspicious. This process allows us to isolate potentially buggy lines without executing the program.

To automate this reasoning, we translate each entailment into a Dafny lemma, where the left-hand side of the entailment becomes the \emph{requires} clause and the right-hand side becomes the \emph{ensures} clause. These lemmas are then submitted to the Dafny verifier, which confirms or rejects them. %

\begin{lstlisting}[label={lst:hoare},caption={Hoare Logic Rules for Dafny Statements in a Buggy \emph{abs} Implementation}]
method abs(x: int) returns (res: int)
  ensures x >= 0 ==> res == x
  ensures x < 0 ==> res == -x
{
  if (x >= 0) {
    // x >= 0
    return x;
    // x >= 0 && res == x (final state)
    // proving post-conditions:
    // (x >= 0 && res == x) ==> (x >= 0 ==> res == x)
    // (x >= 0 && res == x) ==> (x < 0 ==> res == -x)
  } else {
    // x < 0
    return x*1;
    // x < 0 && res == x*1 (final state)
    // proving post-conditions:
    // (x < 0 && res == x*1) ==> (x >= 0 ==> res == x)
    // (x < 0 && res == x*1) ==> (x < 0 ==> res == -x)
  }
}
\end{lstlisting}

We illustrate this process in the comments of~\Cref{lst:hoare}, which involves an \emph{if-then-else} construct for computing  the absolute value of an integer (in this case, \emph{incorrectly}). Entailment checks at each return point reveal whether the final state aligns with the specification. It is important to note that the post-condition does not necessarily include a definition of the final state as a function of the input, as in the example. If any entailment fails, the corresponding code block is flagged. Thus, our tool marks line 14 in the example as a suspicious line, as the implication in line 18 does not hold. Note that we only use Hoare logic rules to reason about the program, thus not relying on symbolic execution.

Using Dafny's verifier directly offers several advantages: it eliminates the need to encode SMT queries manually, avoids duplication of verification logic, and uses native support for Dafny-specific constructs such as sequences and mathematical expressions. However, when translating variable states, we must take care to ensure consistency with the verifier's internal representation. 

This formal fault localization step outputs a ranked list of suspicious lines, which are then passed to the patch generation phase.

\subsection{Patch Generation and Validation}

We generate prompts for each suspicious line, which include the program's context and specification. These prompts are passed to an LLM to produce candidate repairs. We evaluate four models: GPT-4o mini, via API, and local versions of Llama 3, Mistral 7B, and Llemma 7B. To repair~\Cref{lst:hoare}, we mark line 14 as buggy and ask the LLM for a fix. In~\Cref{lst:llm_answer}, we provide a snippet of an answer from Llama 3 that indicated a correct patch: \texttt{return -x;}.

\begin{lstlisting}[language={json}, label={lst:llm_answer},caption={Example Snippet of a Llama 3 Model's Answer}]
"id": "chatcmpl-kvpfrxxhl3hmnbyzd4skm",
"choices": [{
  "index": 0,
  "message": { "content": "return -x;" },
}]
\end{lstlisting}

Our selection of LLMs was based on availability, code generation capability, and mathematical reasoning relevance. GPT-4o mini was chosen for its prior performance in some Dafny-related tasks~\cite{poesia_dafny-annotator_2024}, while Llama 3 and Mistral 7B were selected for their recent release and code synthesis capabilities. Llemma 7B was included for its specialization in mathematical tasks, aligning with formal verification needs.

After the patch generation, we directly insert it into the source code for validation using the Dafny verifier. During the validation, three things can happen:
\begin{itemize}
    \item \textbf{Successful verification}: which means that all specifications hold. In this case, we save the patch in a list of validated patches and continue iterating over the suspicious lines;
    \item \textbf{Failed verification}: this can be broken into the following two cases. \begin{itemize}
        \item \emph{Less than Three Attempts}: In this case, we return to prompt generation to attempt to obtain a better answer from the LLM;
        \item \emph{Third Attempt}: As a limit, we query each line at most three times per model. If verification fails on the third attempt, we skip the suspicious line and continue iterating over the remaining lines.
    \end{itemize}
\end{itemize}

At the end of the iteration, we should have a list of valid patches. If this list is empty, then we deem the fault unable to be fixed. For our example,~\Cref{lst:llm_answer} provided a patch that Dafny verifies, hence it is part of the output list of patches.

\section{Implementation}\label{sec:impl}
Our solution is implemented in C\# and extends the Dafny framework. The tool uses Dafny's existing abstract syntax tree (AST) and built-in verifier to perform fault localization and automated program repair. %
The core implementation is structured around two main components: \texttt{FaultLocalization} and \texttt{Repair}.

\subsection{Fault Localization}

The \texttt{FaultLocalization} component analyzes failed program members by computing the logical state at each statement. Based on Hoare logic rules, it constructs entailments that represent the expected relationship between the pre- and post-states. These entailments are encoded as Dafny lemmas and verified using the built-in Z3-based verifier.

Each program statement is represented by a \texttt{StatementContext}, which tracks a list of \texttt{StateCondition} instances\,---\,logical expressions capturing the evolving program state. A failed entailment causes its \texttt{StateCondition} to be marked as unverified. If any condition associated with a statement fails verification (excluding known valid pre-conditions), the statement is flagged as suspicious.

Different Dafny statement types are mapped to Hoare rules to support this, as shown in the comments in~\Cref{lst:hoare}. For example, \texttt{IfStmt} branches create distinct states for true and false conditions, \texttt{WhileStmt} requires verification of initialization, maintenance, and termination entailments, and \texttt{UpdateStmt} tracks variable changes across assignments.

\subsection{Repair}

Once suspicious lines are identified through fault localization, we attempt to repair them using LLMs via prompt-based generation. Because the models we use are not explicitly trained on Dafny, prompt engineering is essential to guide the models toward generating correct and syntactically valid fixes.

We iterate through the list of suspicious lines and, for each one, we send a carefully constructed prompt to the LLM. The prompt includes the method context and highlights the buggy line with a \texttt{//buggy line} comment. The model is asked to return only the corrected version of that line. Each generated fix is inserted into the original program and verified using the Dafny verifier. As mentioned before, if the program fails verification, the model is queried again, up to three times per line. If no attempt succeeds, the repair process moves to the following suspicious line.

\subsubsection{\bf Prompt Design.}

The content of the prompt is crucial for obtaining a high-quality response from the model~\cite{zhao_explainability_2024}. We structured the prompt as follows:

\begin{enumerate}
    \item \textbf{Role Designation}: Defines the model's role with the instruction: ``You are a software expert specializing in formal methods using the Dafny programming language'';
    \item \textbf{Context Description}: Provides the context and background to the model: ``You receive the following program, where a verifier error message indicates an issue. The error is due to a buggy line marked with the comment `//buggy line.' '';
    \item \textbf{Task Description}: States the objective for the model: ``Your task is to correct the buggy line to ensure the program verifies successfully.'';
    \item \textbf{Buggy Line}: Presents the code with the buggy line marked to inform the model that the bug is in this specific method and line;
    \item \textbf{Ask for Output}: Requests the model to complete the correction by providing the fixed line: ``\textbackslash n fixed line: \textbackslash n''.
\end{enumerate}

We use two distinct roles: the \emph{system role} and the \emph{user role}. The \emph{system role} outlines the model's context, specifying that we want it to act as a software expert in formal verification. The \emph{user role}, on the other hand, involves defining the queries we pose to the model. In our case, we expect the model to return a corrected line of Dafny code.
Listing~\ref{lst:prompt_1} illustrates the first prompt created. 

We conducted several initial experiments with multiple prompt formats using \emph{LM Studio} with the model \textit{Mistral 7B}. We discovered several challenges:

\begin{itemize}
    \item Outputs often included excessive explanation or justification;
    \item Models sometimes returned full methods or wrapped results in inconsistent formats (e.g., using triple backticks, quotes, or language annotations like \texttt{``c''});
    \item Occasionally, models returned syntax-incompatible or misaligned suggestions.
\end{itemize}

\begin{lstlisting}[language={json}, caption={First Prompt for the LLM}, label={lst:prompt_1}]
{   role: system, 
    content: "You are a software expert specializing in formal methods using the Dafny programming language. You receive the following program where a verifier error message indicates an issue. The error is due to a buggy line, which is marked with the comment '//buggy line'. 
    Your task is to correct the buggy line to ensure the program verifies successfully.
    Here is the code: "
},{ role: user,
    content: code + "\nfixed line: \n"
} # code is the original code marked with a buggy line
\end{lstlisting}

After these experiments, we evolved the prompt until we reached a clean output with the desired content. Therefore, as shown in Listing \ref{lst:prompt}, we revised the final prompt to explicitly instruct the model not to include justifications and to return only the fixed line. 
This configuration yielded consistent and usable results, particularly with the Mistral 7B model running locally. To further control cost and runtime, we capped the token output to 30, which is sufficient for our purposes. %
It should be noted, however, that for complex Dafny programs, the model may struggle to produce a fix and might instead provide explanations of how the code could be modified.

\begin{lstlisting}[language={json}, caption={Final Prompt for the LLM}, label={lst:prompt}]
{   role: system, 
    content: "You are a software expert specializing in formal methods using the Dafny programming language. You receive the following program where a verifier error message indicates an issue. The error is due to a buggy line, which is marked with the comment '//buggy line'. 
    Your task is to correct the buggy line to ensure the program verifies successfully.
    Do not include explanations.
    Return only the fixed line.
    Here is the code: "
},{ role: user,
    content: code  + "\nfixed line: \n"
}
\end{lstlisting}

\subsubsection{Tool Execution.}

We implemented our APR tool in C\#
and integrated it with the Dafny verifier. The tool is available on \href{https://github.com/VeriFixer/APRepair-of-Arithmetic-Programs-in-Dafny-using-LLMs}{GitHub}. 
The tool supports both API-based and locally-hosted LLMs. Local models (Llama 3, Mistral 7B, and Llemma 7B) are managed using LM Studio, while GPT-4o mini is accessed via the OpenAI .NET API.

The repair process begins by passing a Dafny program as an argument. If the verifier detects specification violations, the tool triggers our APR components, which perform fault localization, generate prompts for the LLM, and apply candidate fixes. Verification is re-run after each patch attempt. If successful, the tool outputs the corrected line.

If a valid fix is found, the tool prints the line number and the suggested patch, enabling potential integration with IDE plugins like Visual Studio Code.

\section{Evaluation}\label{sec:eval}

In this section, we evaluate the effectiveness of our tool in correctly identifying buggy lines in Dafny programs and generating valid repairs using LLMs. We assess the fault localization and repair success rates across multiple LLMs, using code annotated with specification-based hints to guide patch generation. Additionally, we describe how bugs were systematically introduced, via operator mutations, into the DafnyBench~\cite{loughridge_dafnybench_2024} dataset to create test cases.

All experiments were conducted on a machine running Windows 10 with an Intel Core i7-8565U CPU @ 1.80GHz (up to 1.99GHz) and 16 GB of RAM. The complete source code, configuration files, and instructions for reproducing the results are available in the project's \href{https://github.com/VeriFixer/APRepair-of-Arithmetic-Programs-in-Dafny-using-LLMs}{repository}.%

\subsection{Dataset}

We base our evaluation on DafnyBench~\cite{loughridge_dafnybench_2024}, a benchmark suite consisting of over 750 Dafny programs curated for the purpose of formal software analysis. DafnyBench includes real-world programs and formally annotated examples drawn from multiple sources, making it a robust basis for evaluating fault localization and automated repair in verification-aware settings. The benchmark comprises two main subsets:
\begin{itemize}
    \item \texttt{ground\_truth}: This folder contains original Dafny programs collected from GitHub using the label \texttt{language: Dafny} via the GitHub API, with data gathered up to the end of 2023. It also incorporates examples from Clover~\cite{sun_clover_2024} and dafny-synthesis~\cite{misu_towards_2024} benchmarks;
    \item \texttt{hints\_removed}: This subset is derived from the \texttt{ground\_truth} programs but has key verification hints (e.g., loop invariants, assertions) intentionally removed, while preserving the contracts. The original goal was to evaluate the ability of LLMs to regenerate these missing components using prompt-based synthesis. In our work, we repurpose these simplified or degraded programs to introduce arithmetic bugs and evaluate the robustness of our repair pipeline.
\end{itemize}

We began by filtering DafnyBench programs to identify those that passed verification, ensuring a reliable base for controlled mutation. Arithmetic expressions were then located and systematically mutated to introduce faults. Given that our work is focused on arithmetic bugs, we employed the following four mutation strategies:

\begin{itemize}
    \item Operator Replacement: swapping \texttt{+} and \texttt{-}, \texttt{*} and \texttt{/}, or \texttt{\%} and \texttt{/};
    \item Coefficient Modification: replacing numeric constants $c$ with a random value in the range $[-c, +c]$;
    \item Variable Reordering: shuffling the order of variables in expressions;
    \item Combined Mutation: applying a combination of the above.
\end{itemize}

Each mutation was intended to introduce verification-breaking changes without causing syntax errors. We marked the mutated lines with a \textbf{//buggy line} comment to support evaluation and prompt construction.

The resulting dataset was organized into the following groups:
\begin{itemize}
    \item \texttt{Original\_Code}: The 782 unmodified DafnyBench programs;
    \item \texttt{Correct\_Code}: A subset of programs from \texttt{Original\_Code} that passed verification, comprising 776 from \texttt{ground\_truth} and 230 from \texttt{hints\_removed};
    \item \texttt{Bugs\_Code}: Contains mutated programs, resulting in 2657 mutations based on the \texttt{ground\_truth} programs, and 477 based on the \texttt{hints\_removed}. Those were divided into: \begin{itemize}
        \item \texttt{Hints}: Mutated files annotated with \texttt{//buggy line};
        \item \texttt{Mutations}: Clean versions of the same programs without annotations, used in automated experiments.
    \end{itemize}
\end{itemize}

\subsection{Results and Discussion}

We evaluate our tool's performance on our mutated datasets derived from DafnyBench: \texttt{hints\_removed}, which consists of simplified programs with hints removed, and \texttt{ground\_truth}, which includes more complex and formally annotated Dafny programs. This separation allows us to assess how the complexity of the code affects both fault localization and automated repair using LLMs.

We evaluate the approach using the following metrics:
\begin{itemize}
    \item Repair Success Rate: percentage of programs that were correctly repaired, i.e., passed Dafny verification after patching;
    \item Repair Accuracy: percentage of valid patches that modified the faulty line;
    \item Model Efficiency: average number of attempts per successful repair.
\end{itemize}

Our static localization approach performs well, with most fixes taking just one attempt to fix a line, as can be seen in~\Cref{tab-accuracy}. This shows the effectiveness of Hoare logic-based analysis for identifying arithmetic bugs. Considering the results for both datasets, the list of suspicious buggy lines
produced by our fault localization approach successfully contains the original buggy line in 89.7\% of the programs considered. 

\begin{table}
\centering
\caption{Number of Repair Attempts Needed to Fix the Programs}\label{tab-accuracy}
\begin{tabular}{|c|c|c|c|}
\hline
Dataset	& 1 Attempt & 2 Attempts & 3 Attempts\\
\hline
\texttt{hints\_removed} & 81.09\% (596) & 12.93\% (95) & 5.99\% (44)\\
\texttt{ground\_truth} & 77.40\% (3472) & 15.43\% (692) & 7.18\% (322)\\
\hline
\end{tabular}
\end{table}

GPT-4o mini significantly outperformed other models, successfully repairing over 70\% of programs in both datasets, as shown in~\Cref{tab-models}. Llemma 7B underperformed, likely due to its limited training scope and capacity. GPT-4o mini not only achieves the highest success rate but also does so with fewer attempts per success, making it the most efficient option.

In 95.33\% of successful repairs (averaged across all models), the modification occurred on a line marked as suspicious by the fault localization module, with GPT-4o mini achieving an average of 97.05\%. This validates the synergy between formal fault localization and LLM-based synthesis.

\begin{table}
\centering
\caption{Repair Success Rate and Efficiency by Model for all 447 and 2657 \texttt{hints\_removed} and \texttt{ground\_truth} Mutations}\label{tab-models}
\begin{tabular}{|c|c|c|c|}
\hline
Model & \texttt{hints\_removed} & \texttt{ground\_truth} & Avg. Attempts / Success\\
\hline
GPT-4o mini & 71.59\% (320) & 74.71\% (1985) & 1.14\\
Llama 3 & 46.09\% (206) & 47.27\% (1256) & 1.47\\
Mistral 7B & 42.73\% (191) & 46.41\% (1233) & 1.34\\
Llemma 7B & 4.03\% (18) & 0.49\% (13) & 1.58\\
\hline
\end{tabular}
\end{table}

\paragraph{\bf Evaluation on \texttt{hints\_removed}.}

Out of 782 programs, 230 were verified successfully and were mutated to produce 447 buggy programs. Of the remaining programs, 486 failed verification for various reasons, most commonly unprovable post-conditions (359 cases), while 66 encountered other issues, such as syntax errors or exceeding the 20-second verification time limit. 

Our tool successfully identified the correct buggy line in 397 out of 447 programs (88.8\%). Failures (50 cases) were caused mainly by incorrect parsing of output (e.g., string representations like \texttt{['37\textbackslash n37']} instead of \texttt{[37]}) and entailment mismatches in programs with successive updates to the same variable, where the entailment fails but points to a subsequent update line. As an example, let us consider a program that contains the statements \texttt{s:=1+2; s:=s+1}, where the buggy line is \texttt{s:=1+2} and the correct version is \texttt{s:=1+1}. The state condition that involves the entailment is \texttt{s==1+2+1}, referencing the statement \texttt{s:= s+1}. Therefore, when the entailment fails, the program will only identify the statement \texttt{s:=s+1} as failed. In a repair context, this can be fixed if the model returns \texttt{s:=s} or \texttt{s:=s+1-1}, but, in this situation, the patch is not the same as the original.

On average, about 50\% of lines were flagged as suspicious. For many programs, all lines were flagged, %
especially those with simple sequential statements or minimal specifications. This is aligned with how our fault localization identifies code blocks rather than fine-grained lines (e.g., entire \texttt{else} branches).

GPT-4o mini performed the best in fixing buggy programs, followed by Llama 3 and Mistral 7B. Llemma 7B lagged significantly. The majority of successful repairs occurred on the first attempt, validating our prompt strategy. Llama and Mistral also produced patches closely resembling the original correct lines.

\paragraph{\bf Evaluation on \texttt{ground\_truth}.}

Of the 782 programs, 776 were verified. After excluding the 230 shared with \texttt{hints\_removed}, we introduced arithmetic bugs into 546 remaining programs, yielding 2657 buggy variants.

Correct buggy lines were identified in 2387 out of 2657 cases (89.8\%). The 270 failures stemmed from more nuanced issues:

\begin{itemize}
    \item Termination Reasoning Gaps: Programs with \texttt{while} loops that fail due to ``cannot prove termination'' or ``decreases expression might not decrease'' cannot be correctly analyzed using our partial correctness rules. The corresponding lemmas do not represent the full behaviour of decreasing expressions, leading to missed bug identification;
    \item Incorrect Lemma Validation: Some invalid entailments are incorrectly verified due to limitations in how we encode lemmas, allowing buggy lines to appear correct;
    \item Timeouts: Lemma checks exceeding the 20-second verifier limit are considered failed;
    \item Unsupported Constructs: Programs containing unsupported statements or ghost variables could not be thoroughly analyzed;
    \item Incorrect Lemma Sorting: When there are more than 10 entailments (named \texttt{check\_N}), lexicographical sorting (e.g., \texttt{check\_10} before \texttt{check\_2}) causes incorrect lemma-to-entailment associations. This prevents accurate mapping of failures to code statements. Sorting by numeric index would resolve this, but it was not implemented. %
\end{itemize}

The average coverage of suspicious lines in \texttt{ground\_truth} was about 70\%, with fewer fully-flagged methods (23.82\%) than in \texttt{hints\_removed}, indicating improved precision in more complex programs. Programs with 100\% flagged lines tended to be shorter (5 to 6 lines).

Results mirrored those of \texttt{hints\_removed}. Again, GPT-4o mini outperformed all other models. First-attempt repairs succeeded in 66.15\% of cases. Of the successful patches, 91.34\% modified the correct line and 80.78\% exactly matched the original correct line.

\paragraph{\bf Combined Insights.}

By synthesizing results across both datasets, we derive the following key insights:

\begin{itemize}
    \item {\bf Model Repair Effectiveness:} The success rates of GPT-4o mini, Llama 3, and Mistral 7B remain consistent across datasets, with GPT-4o mini achieving the highest repair rate;
    \item{\bf Repair Despite Incomplete Localization:} In some cases, the correct fix was generated even when the buggy line was not flagged by fault localization. This demonstrates the generative flexibility of LLMs;
    \item {\bf Prompt Efficiency:} Most repairs succeeded on the first attempt, confirming the effectiveness of our prompt design. Low success rates in the second and third attempts highlight cases where the model repeatedly proposed the same incorrect patch.

\end{itemize}

\paragraph{\bf Limitations.}
Despite promising results, several limitations remain:
\begin{itemize}
\item {\bf Specification correctness:} Our method assumes that the formal specification is correct, not necessarily having all the assertions and invariants needed for Dafny to prove it. Bugs can also reside in the specification, but repairing or suggesting improvements to specs is outside the scope of this work; %
\item {\bf Bug scope:} We focused exclusively on arithmetic errors. Our approach does not yet support other error classes (e.g., control flow, heap misuse);
\item {\bf Model hallucination and noise:} While in many cases LLMs generated high-quality patches, they occasionally introduced irrelevant changes or produced code with syntactic errors. We mitigate this with automated validation, but model outputs remain unpredictable;
\item {\bf Prompt sensitivity:} 
Repair quality is sensitive to prompt formulation; subtle changes in wording, context, or formatting can significantly affect outcomes.

\end{itemize}

\paragraph{Summary of Results:}
    Our tool achieves high fault localization success rate (89.7\%) using formal specifications and static analysis alone. GPT-4o mini is the most effective model, repairing 74.18\% of faulty programs with a low average attempt count. Also, combining formal methods for localization with LLM-based synthesis yields repairs that are both correct and efficient. In addition, verification-based validation ensures that accepted patches respect the full program specification.

It should be noted that a potential threat to the validity of our results is that the LLM may have seen the correct code versions during training,  which may positively influence the results. Nonetheless, the buggy versions were created by us, so the model has no prior information about the correspondence between the buggy and correct versions. Furthermore, in the context of Dafny, real-world buggy examples are scarce, so our experimental setup necessarily relies on artificially created examples.

\section{Conclusion}\label{sec:concl}
This work proposes an automated program repair approach tailored for Dafny, targeting arithmetic bugs and using formal specifications as correctness oracles.

We combine formal fault localization with LLM-guided repair using GPT-4o mini, Llama 3, Mistral 7B, and Llemma 7B. 
We achieve a success rate of 89.7\% in identifying the buggy line.
However, limitations arise in programs with repeated updates to the same variable and termination verification issues in \texttt{while} loops due to the use of partial correctness reasoning. These limitations suggest areas for refinement in entailment modelling and verifier integration.

GPT-4o mini was the most effective in generating valid patches. While Llama 3 and Mistral 7B also showed potential, Llemma 7B underperformed, likely due to its poor alignment with the structure and semantics of Dafny. %

Future work includes support for bug types beyond arithmetic errors, including those involving control flow, specification violations, and method contracts. We also plan to develop a Visual Studio Code extension that integrates our tool, enabling real-time bug repair suggestions based on formal specifications. %

\begin{credits}
\subsubsection{\ackname} 
This work was financed by National Funds through the FCT - Fundação para a Ciência e a Tecnologia, I.P. (Portuguese Foundation for Science and Technology) within the project VeriFixer, with reference 2023.15557.PEX (DOI: 10.54499/2023.15557.PEX). 
Alexandre Abreu was financed by National Funds through the Portuguese funding agency, FCT, within the project LA/P/0063/2020 (DOI: 10.\allowbreak54499/LA/P/0063/2020) and grant 2024.00375.BD.

\end{credits}

\bibliographystyle{splncs04}
\bibliography{bibliography}

\begin{thebibliography}{10}
\providecommand{\url}[1]{\texttt{#1}}
\providecommand{\urlprefix}{URL }
\providecommand{\doi}[1]{https://doi.org/#1}

\bibitem{abreu_exploring_2023}
Abreu, A., Macedo, N., Mendes, A.: Exploring {Automatic} {Specification} {Repair} in {Dafny} {Programs}. In: 2023 {38th} {IEEE}/{ACM} {Int.} {Conference} {on} {Automated} {Software} {Engineering} {Workshops}, {ASEW}. pp. 105--112. {IEEE} {ACM} {International} {Conference} on {Automated} {Software} {Engineering}, IEEE COMPUTER SOC (2023)

\bibitem{abreu_accuracy_2007}
Abreu, R., Zoeteweij, P., van Gemund, A.J.: On the {Accuracy} of {Spectrum}-based {Fault} {Localization}. In: Testing: {Academic} and {Industrial} {Conference} {Practice} and {Research} {Techniques} - {MUTATION}. pp. 89--98 (Sep 2007)

\bibitem{carreira2025DafnyStudentStudy}
Carreira, C., Silva, {\'A}., Abreu, A., Mendes, A.: Can large language models help students prove software correctness? {An} experimental study with {D}afny. In: 23rd Int. Conf. on Software Engineering and Formal Methods (SEFM) (2025)

\bibitem{feng_codebert_2020}
Feng, Z., Guo, D., Tang, D., Duan, N., Feng, X., Gong, M., Shou, L., Qin, B., Liu, T., Jiang, D., Zhou, M.: {CodeBERT}: {A} {Pre}-{Trained} {Model} for {Programming} and {Natural} {Languages} (Sep 2020)

\bibitem{hoare_axiomatic_1969}
Hoare, C.A.R.: An axiomatic basis for computer programming. Commun. ACM  \textbf{12}(10),  576--580 (Oct 1969)

\bibitem{konighofer_automated_2011}
Könighofer, R., Bloem, R.: Automated error localization and correction for imperative programs. In: Proceedings of the {International} {Conference} on {Formal} {Methods} in {Computer}-{Aided} {Design}. pp. 91--100. FMCAD Inc, Austin, Texas (Oct 2011)

\bibitem{le_goues_genprog_2012}
Le~Goues, C., Nguyen, T., Forrest, S., Weimer, W.: {GenProg}: {A} {Generic} {Method} for {Automatic} {Software} {Repair}. IEEE Transactions on Software Engineering  \textbf{38}(1),  54--72 (Jan 2012)

\bibitem{le_goues_automated_2019}
Le~Goues, C., Pradel, M., Roychoudhury, A.: Automated program repair. Commun. ACM  \textbf{62}(12),  56--65 (Nov 2019)

\bibitem{le_goues_automatic_2021}
Le~Goues, C., Pradel, M., Roychoudhury, A., Chandra, S.: Automatic {Program} {Repair}. IEEE Software  \textbf{38}(4),  22--27 (Jul 2021)

\bibitem{leino2010dafny}
Leino, K.R.M.: Dafny: An automatic program verifier for functional correctness. In: Int. Conf. on Logic for Programming Artificial Intelligence and Reasoning. pp. 348--370. Springer (2010)

\bibitem{leino_dafny_2014}
Leino, K.R.M., Wüstholz, V.: The {Dafny} {Integrated} {Development} {Environment}. Electron. Proc. Theor. Comput. Sci.  \textbf{149},  3--15 (Apr 2014)

\bibitem{loughridge_dafnybench_2024}
Loughridge, C., Sun, Q., Ahrenbach, S., Cassano, F., Sun, C., Sheng, Y., Mudide, A., Misu, M.R.H., Amin, N., Tegmark, M.: Dafnybench: A benchmark for formal software verification. arXiv preprint arXiv:2406.08467  (2024)

\bibitem{designbycontract}
Meyer, B.: Design by contract. Prentice Hall Upper Saddle River (2002)

\bibitem{misu_towards_2024}
Misu, M.R.H., Lopes, C.V., Ma, I., Noble, J.: Towards {AI}-{Assisted} {Synthesis} of {Verified} {Dafny} {Methods}. Proc. ACM Softw. Eng.  \textbf{1}(FSE),  812--835 (Jul 2024)

\bibitem{apr_17}
Monperrus, M.: The living review on automated program repair. Technical Report hal-01956501, {HAL Archives Ouvertes} (2018)

\bibitem{de_moura_z3_2008}
de~Moura, L., Bjørner, N.: Z3: {An} {Efficient} {SMT} {Solver}. In: Ramakrishnan, C.R., Rehof, J. (eds.) Tools and {Algorithms} for the {Construction} and {Analysis} of {Systems}. pp. 337--340. Springer, Berlin, Heidelberg (2008)

\bibitem{moy_testing_2013}
Moy, Y., Ledinot, E., Delseny, H., Wiels, V., Monate, B.: Testing or {Formal} {Verification}: {DO}-{178C} {Alternatives} and {Industrial} {Experience}. IEEE Software  \textbf{30}(3),  50--57 (May 2013)

\bibitem{mugnier_laurel_2024}
Mugnier, E., Gonzalez, E.A., Polikarpova, N., Jhala, R., Yuanyuan, Z.: Laurel: Unblocking automated verification with large language models. Proc. ACM Program. Lang.  \textbf{9}(OOPSLA1) (Apr 2025)

\bibitem{llms}
Naveed, H., Khan, A.U., Qiu, S., Saqib, M., Anwar, S., Usman, M., Akhtar, N., Barnes, N., Mian, A.: A comprehensive overview of {Large Language Models}. ACM Trans. Intell. Syst. Technol.  (Jun 2025)

\bibitem{nguyen_semfix_2013}
Nguyen, H.D.T., Qi, D., Roychoudhury, A., Chandra, S.: {SemFix}: {Program} repair via semantic analysis. In: 2013 35th {International} {Conference} on {Software} {Engineering} ({ICSE}). pp. 772--781 (May 2013)

\bibitem{nguyen_automatic_2019}
Nguyen, T.T., Ta, Q.T., Chin, W.N.: Automatic {Program} {Repair} {Using} {Formal} {Verification} and {Expression} {Templates}. In: Enea, C., Piskac, R. (eds.) Verification, {Model} {Checking}, and {Abstract} {Interpretation}. pp. 70--91. Springer International Publishing, Cham (2019)

\bibitem{pei_automatic_2014}
Pei, Y., Furia, C.A., Nordio, M., Meyer, B.: Automatic {Program} {Repair} by {Fixing} {Contracts}. In: Gnesi, S., Rensink, A. (eds.) Fundamental {Approaches} to {Software} {Engineering}. pp. 246--260. Springer, Berlin, Heidelberg (2014)

\bibitem{pei_code-based_2011}
Pei, Y., Wei, Y., Furia, C.A., Nordio, M., Meyer, B.: Code-based automated program fixing. In: 2011 26th {IEEE}/{ACM} {International} {Conference} on {Automated} {Software} {Engineering} ({ASE} 2011). pp. 392--395 (Nov 2011)

\bibitem{poesia_dafny-annotator_2024}
Poesia, G., Loughridge, C., Amin, N.: dafny-annotator: {AI}-{Assisted} {Verification} of {Dafny} {Programs} (Nov 2024)

\bibitem{prenner_can_2022}
Prenner, J.A., Babii, H., Robbes, R.: Can {OpenAI}'s codex fix bugs? an evaluation on {QuixBugs}. In: Proc. of the {Third} {International} {Workshop} on {Automated} {Program} {Repair}. pp. 69--75. {APR} '22, ACM, New York, NY, USA (Oct 2022)

\bibitem{silva_leveraging_2024}
Silva, A.F., Mendes, A., Ferreira, J.F.: Leveraging {Large} {Language} {Models} to {Boost} {Dafny}’s {Developers} {Productivity}. In: Proceedings of the 2024 {IEEE}/{ACM} 12th {Int.} {Conference} on {Formal} {Methods} in {Software} {Engineering}. pp. 138--142. {FormaliSE} '24, Association for Computing Machinery, New York, NY, USA (Jun 2024)

\bibitem{smith_is_2015}
Smith, E.K., Barr, E.T., Le~Goues, C., Brun, Y.: Is the cure worse than the disease? overfitting in automated program repair. In: Proceedings of the 2015 10th {Joint} {Meeting} on {Foundations} of {Software} {Engineering}. pp. 532--543. {ESEC}/{FSE} 2015, Association for Computing Machinery, New York, NY, USA (Aug 2015)

\bibitem{sun_clover_2024}
Sun, C., Sheng, Y., Padon, O., Barrett, C.: Clover: {Closed}-{Loop} {Verifiable} {Code} {Generation}. In: Avni, G., Giacobbe, M., Johnson, T.T., Katz, G., Lukina, A., Narodytska, N., Schilling, C. (eds.) {AI} {Verification}. pp. 134--155. Springer Nature Switzerland, Cham (2024)

\bibitem{wei_automated_2010}
Wei, Y., Pei, Y., Furia, C.A., Silva, L.S., Buchholz, S., Meyer, B., Zeller, A.: Automated fixing of programs with contracts. In: Proceedings of the 19th international symposium on {Software} testing and analysis. pp. 61--72. {ISSTA} '10, Association for Computing Machinery, New York, NY, USA (Jul 2010)

\bibitem{wong_dstar_2014}
Wong, W.E., Debroy, V., Gao, R., Li, Y.: The {DStar} {Method} for {Effective} {Software} {Fault} {Localization}. IEEE Transactions on Reliability  \textbf{63}(1),  290--308 (Mar 2014)

\bibitem{wong_survey_2016}
Wong, W.E., Gao, R., Li, Y., Abreu, R., Wotawa, F.: A {Survey} on {Software} {Fault} {Localization}. IEEE Trans. on Software Engineering  \textbf{42}(8),  707--740 (Aug 2016)

\bibitem{xia_automated_2023}
Xia, C.S., Wei, Y., Zhang, L.: Automated {Program} {Repair} in the {Era} of {Large} {Pre}-trained {Language} {Models}. In: 2023 {IEEE}/{ACM} 45th {International} {Conference} on {Software} {Engineering} ({ICSE}). pp. 1482--1494 (May 2023)

\bibitem{xia_less_2022}
Xia, C.S., Zhang, L.: Less training, more repairing please: revisiting automated program repair via zero-shot learning. In: Proc. of the 30th {ACM} {Joint} {European} {Software} {Engineering} {Conf.} and {Symposium} on the {Foundations} of {Software} {Engineering}. pp. 959--971. {ESEC}/{FSE} 2022, ACM, New York, NY, USA (Nov 2022)

\bibitem{zhao_explainability_2024}
Zhao, H., Chen, H., Yang, F., Liu, N., Deng, H., Cai, H., Wang, S., Yin, D., Du, M.: Explainability for {Large} {Language} {Models}: {A} {Survey}. ACM Trans. Intell. Syst. Technol.  \textbf{15}(2),  20:1--20:38 (Feb 2024)

\end{thebibliography}
\end{document}